\documentclass[10pt]{article} \usepackage{aaspp4}

\newcommand{\HI}{H$\;${\small\rm I}\relax}

\newcommand{\AlIII}{Al$\;${\small\rm III}\relax}
\newcommand{\Ha}{H$\alpha$}
\newcommand{\lya}{Ly$\alpha$}
\newcommand{\dlas}{DLAs}
\newcommand{\z}{$z$}

\newcommand{\fhii}{$f({\rm H^+})$}

\newcommand{\wave}[1]{$\lambda$#1\relax}

\newcommand{\cloudy}{CLOUDY}
\newcommand{\etal}{{\em et al.}\relax}

\newcommand{\msun}{M$_\odot$}

\newcommand{\column}{cm$^{-2}$}

\newcommand{\e}[1]{10^{#1}}

\begin{document}

\title{Ionized Gas in Damped Lyman-$\alpha$ Systems and
             Its Effects on \\Elemental Abundance Studies}

\author{J. Christopher Howk \& Kenneth R. Sembach}
\affil{Department of Physics \& Astronomy, The Johns Hopkins University,\\
3400 N. Charles St., Baltimore, MD  21218 \\
{\it howk@pha.jhu.edu, sembach@pha.jhu.edu}}

\authoremail{howk@pha.jhu.edu}

\begin{center}
To appear in {\em The Astrophysical Journal Letters.}
\end{center}

\begin{abstract}

Recent high-resolution observations of metal absorption lines in
high-redshift damped \lya\ systems have shown that \ion{Al}{3}, a
tracer of moderately-ionized gas, very often has a velocity structure
indistinguishable from that of low-ionization gas.  Regions of ionized
and neutral hydrogen in these systems are likely cospatial.  The
higher-ionization \ion{Si}{4} and \ion{C}{4} absorption shows a much
weaker or non-existent correlation with the low ionization material,
implying that the regions traced by \ion{Al}{3} are photoionized by a
soft (stellar) spectrum, by a hard (power law) spectrum with a very
low ionization parameter, or a combination of both.  We discuss the
ionization of the damped \lya\ systems and use photoionization
equilibrium models to make quantitative estimates of its effects on
abundance studies in these systems.  We show that ionization effects
may be large enough to account for the observed dispersion in absolute
metal abundances in damped \lya\ systems, causing systematically
higher abundances in lower column density systems.  The observed
Si$^+$/Fe$^+$ and Zn$^+$/Cr$^+$ ratios may systematically overestimate
the intrinsic Si/Fe and Zn/Cr ratios, respectively, if ionized gas is
present in these systems, thereby mimicking the effects of
$\alpha$-element enrichment or dust depletion.

\end{abstract}



\section{EVIDENCE FOR PHOTOIONIZED GAS IN DAMPED Ly$\alpha$ SYSTEMS}

Observations of damped \lya\ systems (\dlas), which may represent the
progenitors to modern disk galaxies, along the sightlines to
high-redshift QSOs allow astronomers to trace the evolution of
elemental abundances over 90\% of the age of the Universe.  This is
typically achieved by comparing the measured column density of a
single ion of a given element, $X^i$, with that of neutral hydrogen,
${\rm H^o}$.  The assumption is then made that the unobserved
ionization stages make negligible contributions and $N(X^i)/N({\rm
H^o}) \approx N(X)/N({\rm H})$.  

The total \HI\ column densities of \dlas\ are, by definition,
$N(\mbox{\HI}) \ga 2\times \e{20}$ \column; if collected in a single
cloud, such a large column density would imply small ionization
corrections.  Although some of these systems include such monolithic
absorbers, high-resolution data
show \dlas\ are often made up of a collection of several (in many
cases $\ga 5 - 10$) lower column density clouds (Lu \etal\ 1996;
Pettini \etal\ 1999; Prochaska \& Wolfe 1999).  Furthermore, both the
Lu \etal\ (1996) and Prochaska \& Wolfe (1999) surveys of
\dlas\ find a very conspicuous correlation between the velocity
structure seen in the absorption lines of low-ionization species
(e.g., \ion{Si}{2}, \ion{Fe}{2}, and \ion{Zn}{2}) and the structure
observed in \AlIII, a tracer of moderately (photo)ionized gas
[IP$({\rm Al}^+,{\rm Al}^{+2})=(18.8,28.4)$ eV].  Such an obvious
correlation is not observed between the low-ionization species and the
more highly-ionized ions such as \ion{Si}{4} or \ion{C}{4}.

Similar arrangements of low- and intermediate-ions can be found along
selected sightlines extending into the halo of the Milky Way.  Towards
HD 93521 (Spitzer \& Fitzpatrick 1993) and $\rho$~Leo (Howk \& Savage
1999) the tracers of neutral and photoionized gas have relative
velocity component distributions resembling those of \ion{Al}{3} and
low ions in \dlas.  The total hydrogen column densities towards these
stars are $\log N(\mbox{\HI}) = 20.10$ and 20.44, respectively (Diplas
\& Savage 1994).  In the Milky Way, the scale height of \ion{Al}{3} 
is consistent with that of the free electrons, $h_z \sim 1$ kpc, which
is more extended than the \HI\ distribution (Savage, Edgar, \& Diplas
1990).  

Most singly-ionized metal species that are dominant ionization stages
in \HI -bearing regions may also be produced in photoionized clouds
where H$^{\rm o}$ is a small fraction of the total hydrogen content.
The formation of metal absorption lines in both ionized and neutral
regions can have a significant effect on elemental abundance
determinations.  Ionization can be an important issue for
high-precision studies of elemental abundances in the Milky Way (Sofia
\& Jenkins 1998; Howk, Savage, \& Fabian 1999; Sembach \etal\ 1999).  
The \AlIII\ in \dlas, with velocity structure that is often
indistinguishable from that of the low ions (Lauroesch \etal\ 1996),
suggests the long-held assumption that ionization effects are neglible
in these systems may be unwarranted.  In this work we examine the
contribution of photoionized gas to the observed metal-line absorption
in damped \lya\ systems.

\section{THE IONIZING SPECTRUM}
\label{sec:spectrum}

The major uncertainty in determining the ionization balance in the
\dlas\ is the unknown shape of the ionizing spectrum.  The two most
likely origins for ionizing photons in \dlas\ are: internal stellar
and external background sources.  Ionization of the \dlas\ by external
sources, e.g., by the integrated light from QSOs, AGNs, starbursts,
and normal galaxies (Haardt \& Madau 1996; Madau \& Shull 1996),
requires that the ionizing photons ``leak'' into the \dlas.  This
might seem unlikely given the large observed neutral hydrogen column
densities, but the multi-component nature of these systems implies
that each individual cloud may have a much lower column density than
the total.  Furthermore, the ionization of the warm ionized medium
(WIM) in the Milky Way requires $\sim15\%$ of the ionizing photon
output of Galactic OB stars (Reynolds 1993).  This implies that the
gaseous structure of a present day disk galaxy is such that ionizing
photons can travel very large distances from their origin, and of
order $5\%$ may escape the Galaxy completely (Bland-Hawthorn \&
Maloney 1999).  We assume a similar arrangement in the \dlas.  For the
external ionization case, we adopt an updated version of the Haardt \&
Madau (1996; hereafter HM) QSO ultraviolet background in our
photoionization models.  This modified background spectrum (Haardt
1999, priv. comm.)  assumes $q_o = 0.5$ (instead of 0.1), a power law
index for the QSO emission spectrum of $\alpha = 1.8$ (rather than
1.5), and a redshift evolution of the QSO number density that follows
the trend described by Madau, Haardt, \& Rees (1999).

Internal ionization, in this work, refers to photoionization by
stellar sources internal to the \dlas.  If \dlas\ represent the early
phases of massive disk galaxies (e.g., Wolfe \& Prochaska 1998), it is
reasonable to expect some star formation in these systems.  Searches
for \lya\ and \Ha\ emission from \dlas\ imply low star formation
rates: $\mbox{\.{M}}_* \la 5 - 20$ \msun\ yr$^{-1}$ (Bunker \etal\
1999; Lowenthal \etal\ 1995), with one detection of \lya\ emission
suggesting $\mbox{\.{M}}_* \approx 1$~\msun\ yr$^{-1}$ (Warren \&
M{\o}ller 1996).  In the Milky Way, where ionizing photons from
early-type stars must leak through the neutral ISM to ionize the WIM,
the star formation rate is of order $2-5$~\msun\ yr$^{-1}$ (Mezger
1987; McKee 1989; McKee \& Williams 1997).  The perpendicular column
density of ionized hydrogen in the WIM is about 1/4 that of neutral
hydrogen at the solar circle, thus demonstrating that a relatively
large fraction of interstellar hydrogen can be ionized with a modest
level of star formation.  For the internal ionization case, we adopt
the spectrum of a typical late O star as the ionizing spectrum.  We
use an ATLAS line-blanketed model atmosphere (Kurucz 1991) with an
effective temperature $T_{eff} =33,000$ K and $\log (g) = 4.0$.  Our
work on the ionization of the Galactic WIM (Sembach \etal\ 1999)
suggests that such a spectrum is able to match the constraints imposed
by emission line observations of the ionized gas (Reynolds \& Tufte
1995; Reynolds \etal\ 1998; Haffner, Reynolds, \& Tufte 1999).

We consider only a single temperature stellar source for the internal
case, and a QSO-dominated spectrum for the external ionization case.
The reader should be aware that the true ionizing spectrum may be a
combination of soft (internal) and hard (external) ionizing spectra.
The lack of associated \ion{Si}{4} absorption with the low ions favors
either the softer stellar spectrum or a very low ionization parameter.

\section{PHOTOIONIZATION MODELS}
\label{sec:models}

We use the \cloudy\ ionization equilibrium code (Ferland \etal\ 1998;
Ferland 1996) to model the ionization of \dlas.  We assume a
plane-parallel geometry with the ionizing spectrum incident on one
side.  Rather than match the total \HI\ column density in our models,
we stop the integration at the point where the local ionization
fraction of neutral hydrogen climbs above 10\%, i.e., 
$x({\rm H^o})\equiv N({\rm H^o})/N({\rm H_{tot}}) > 0.1$.  Our models
therefore treat the (almost) fully-ionized regions assumed to envelop
the neutral, \HI -bearing clouds.  The relative mix of neutral and
ionized material can be inferred from observations of adjacent ions,
e.g., \ion{Al}{2}/\AlIII.  Our models assume a base metal abundance of
0.1 solar, with relative heavy element abundances equivalent to those
observed in the Galactic warm neutral medium (Sembach \etal\ 1999;
Howk \etal\ 1999).  We include interstellar grains for heating and
cooling processes (see Ferland 1996 and Baldwin \etal\ 1991), with a
dust to gas ratio 0.1 of the Galactic value.  Our models are only as
accurate as the input atomic data for the \cloudy\ code, and we refer
the reader to Ferland (1996) and Ferland \etal\ (1998) for discussions
of the uncertainties (see also our earlier work with \cloudy: Sembach
\etal\ 1999; Howk \& Savage 1999).  In particular, the dielectronic
recombination coefficients for elements in the third and fourth row of
the periodic table are typically not well known, and the radiative
recombination coefficients for many of the heavier elements (e.g., Zn
and Cr) are often based on somewhat uncertain theoretical
considerations.

We have computed \cloudy\ models for the $z \approx 2.0$ HM spectrum
and for the Kurucz model atmosphere over a range of ionization
parameters, $\Gamma$.  In this case $\Gamma$ is the dimensionless
ratio of total hydrogen-ionizing photon density to hydrogen particle
density at the face of the cloud.  In Figure \ref{fig:hmfracs} we
present the ionization fractions of several ions, $x(X^i)$, for the HM
spectrum as a function of the assumed ionization parameter.  The top
panel shows $x(X^i)$ for elements with at least two potentially
measurable ionization stages: Si, Fe, and Al.  The bottom panel shows
the effects of ionization on relative metal abundances, tracing values
of $x(X^i)/x({\rm Fe}^+)$ for several commonly measured ions.  These
plots can be used to correct for ionization effects if one is able to
estimate $\Gamma$.

For large values of $\log \Gamma \ (\ga -3.0)$ the predicted strength
of the \ion{Si}{4} becomes large, with $x({\rm Si^+})/x({\rm Si^{+3}})
\la 2$, contrary to observations.  At $\log \Gamma = -4.0$ this
ratio is $\approx 100$.  We note that the behavior of the ratio
$x({\rm Ni^+})/x({\rm Cr^+})$ in Figure
\ref{fig:hmfracs} also suggests a low ionization fraction, given that the
observed ratio $N({\rm Ni^+})/N({\rm Cr^+})$ is typically very near
the solar Ni/Cr ratio.\footnote{This result relies on new $f$-value
determinations by Fedchak \& Lawler (1999).  Using these new
oscillator strengths we find a (weighted) average abundance $[\langle
{\rm Cr/Ni} \rangle] = +0.013\pm0.023$ in the 11 \dlas\ containing
both elements in the Prochaska \& Wolfe (1999) sample.}  The utility
of this ratio as an indicator of the ionization parameter would be
improved with better atomic data.  Figure \ref{fig:hmfracs} shows that
while \AlIII\ is a tracer of ionized gas, it accounts for less than
10\% of the total aluminum column density, even in regions of
fully-ionized hydrogen (where \ion{Al}{2} or \ion{Al}{4} dominate).
Unfortunately, this implies that past arguments for a lack of ionized
gas in \dlas\ based upon a relatively large \ion{Al}{2}/\AlIII\ ratio
are possibly erroneous.

Figure \ref{fig:stellarfracs} shows the \cloudy\ photoionization
calculations performed assuming internal sources of ionizing photons,
i.e., star formation.  Again the fraction of aluminum in
\ion{Al}{3} is relatively small.  If we assume that the properties 
of the ionized gas in the \dlas\ are similar to those of the WIM in
the Milky Way, a relatively low ionization parameter is preferred
(e.g., $\log \Gamma \la -3.7$ is adopted by Sembach \etal\ 1999).  The
$x({\rm Ni^+})/x({\rm Cr^+})$ ratio suggests a low value of $\Gamma$,
as in the external ionization case.  For the adopted stellar spectrum,
the fraction of silicon in the form of Si$^{+3}$ never rises above
0.1\% for the range of ionization parameters considered.  Note that
this is a considerably smaller fraction than found for high-\z\ Lyman
limit systems (Steidel \& Sargent 1989; Prochaska 1999) and
\lya\ forest clouds (Songaila \& Cowie 1996).

\section{DISCUSSION}
\label{sec:discussion}

\begin{figure}
\epsscale{1.0}
\plotone{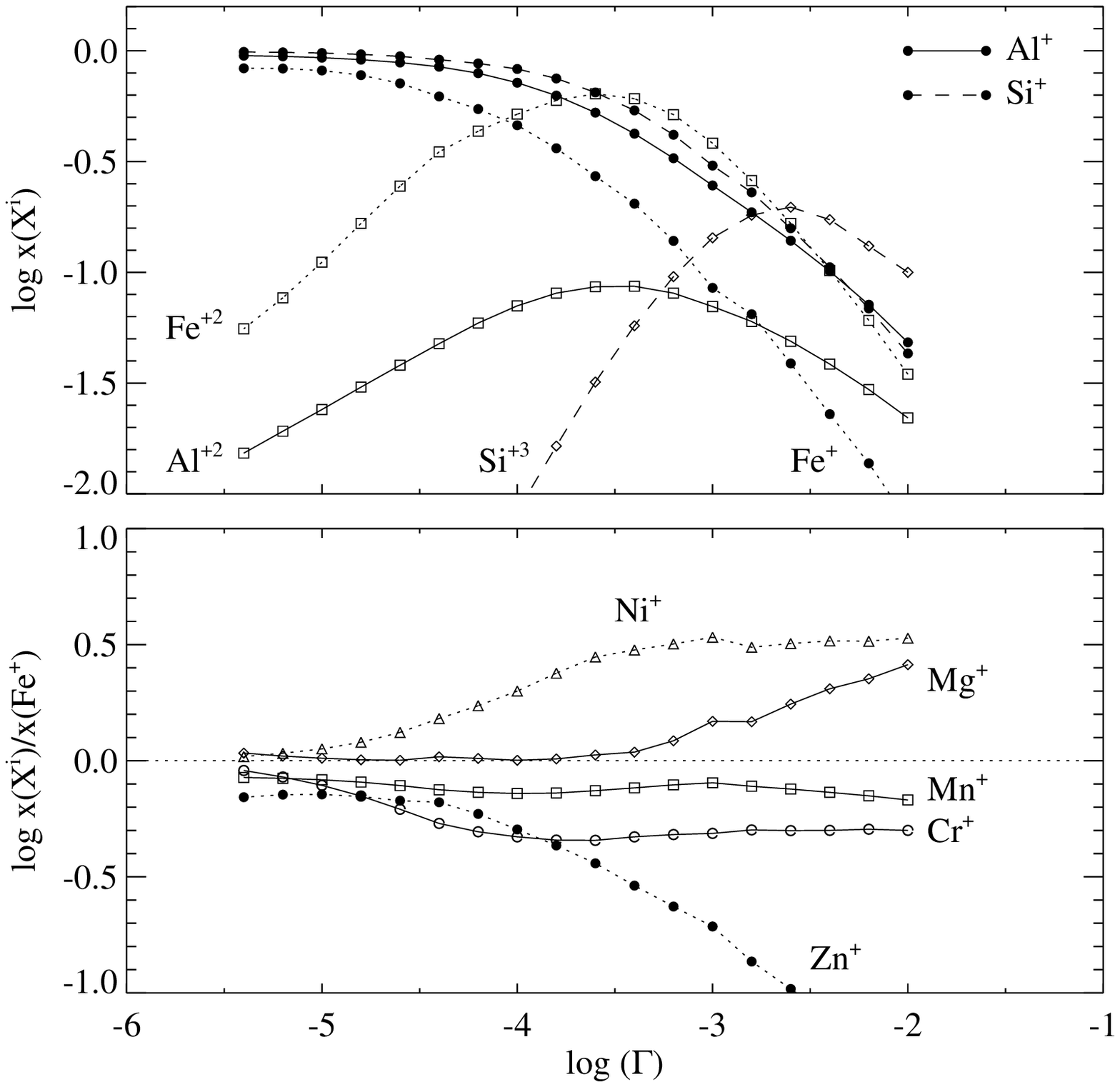}
\caption{Ionization fractions as a function of ionization parameter, 
$\Gamma$, for several ions assuming a Haardt \& Madau (1996) UV
background spectrum appropriate for $z=2.0$.  The top panel shows the
ionization fractions of ions of Fe, Al, and Si, all of which have
multiple potentially observable ionization states.  The bottom panel
shows the ionization fractions of several commonly observed
singly-ionized species relative to that of Fe$^+$.  These ionization
fractions are appropriate for the fully-ionized gas in a system.  The
atomic input data used for Cr and Zn, in particular, are somewhat
uncertain (see text).
\label{fig:hmfracs}}
\end{figure}

\begin{figure}
\epsscale{1.0}
\plotone{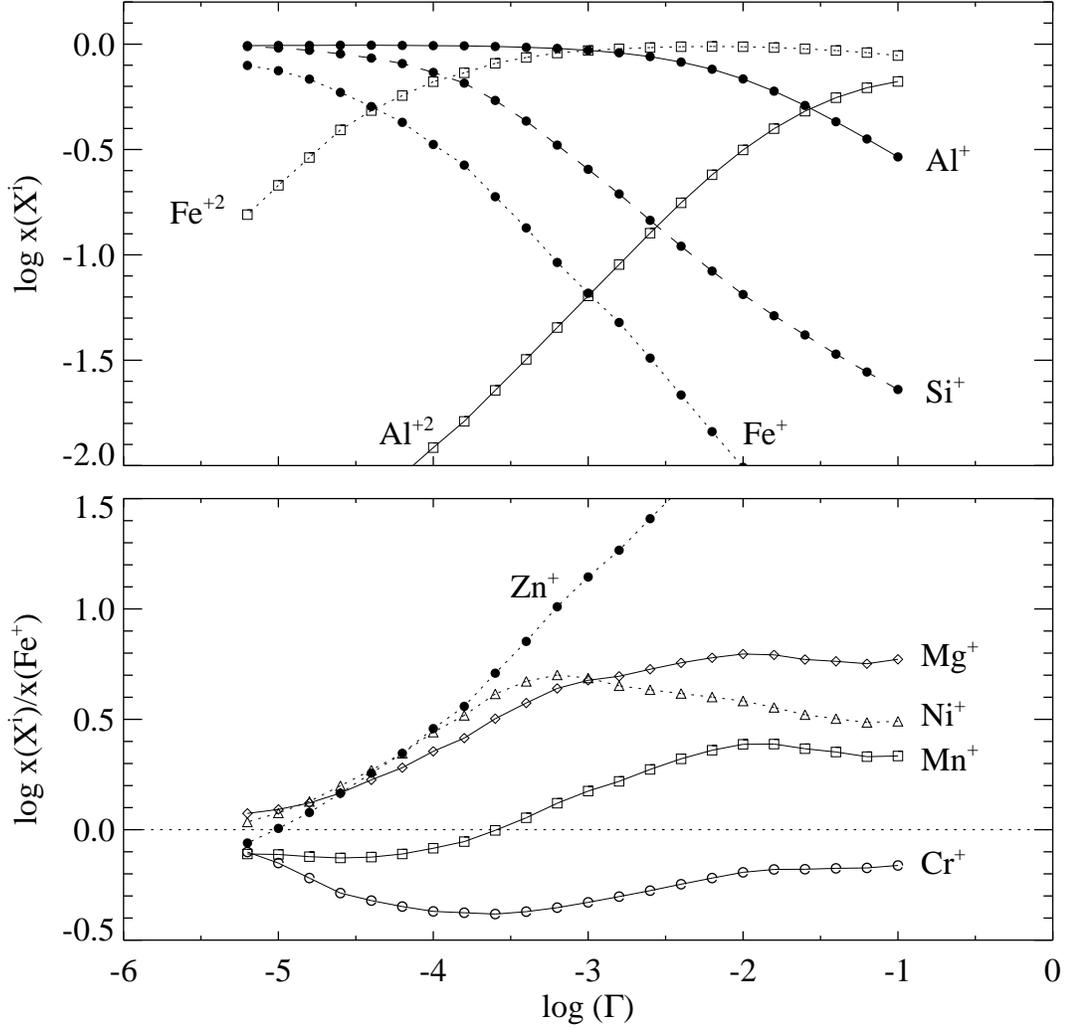}
\caption{
As Figure \ref{fig:hmfracs} but assuming an internal source of
ionization approximated by a Kurucz (1991) model atmosphere with
$T_{eff} = 33,000$ K.  Si$^{+3}$ is not seen on this plot because for
all ionization parameters considered, $\log x({\rm Si}^{+3}) < -3$.
\label{fig:stellarfracs}}
\end{figure}

\begin{figure}
\epsscale{0.7}
\plotone{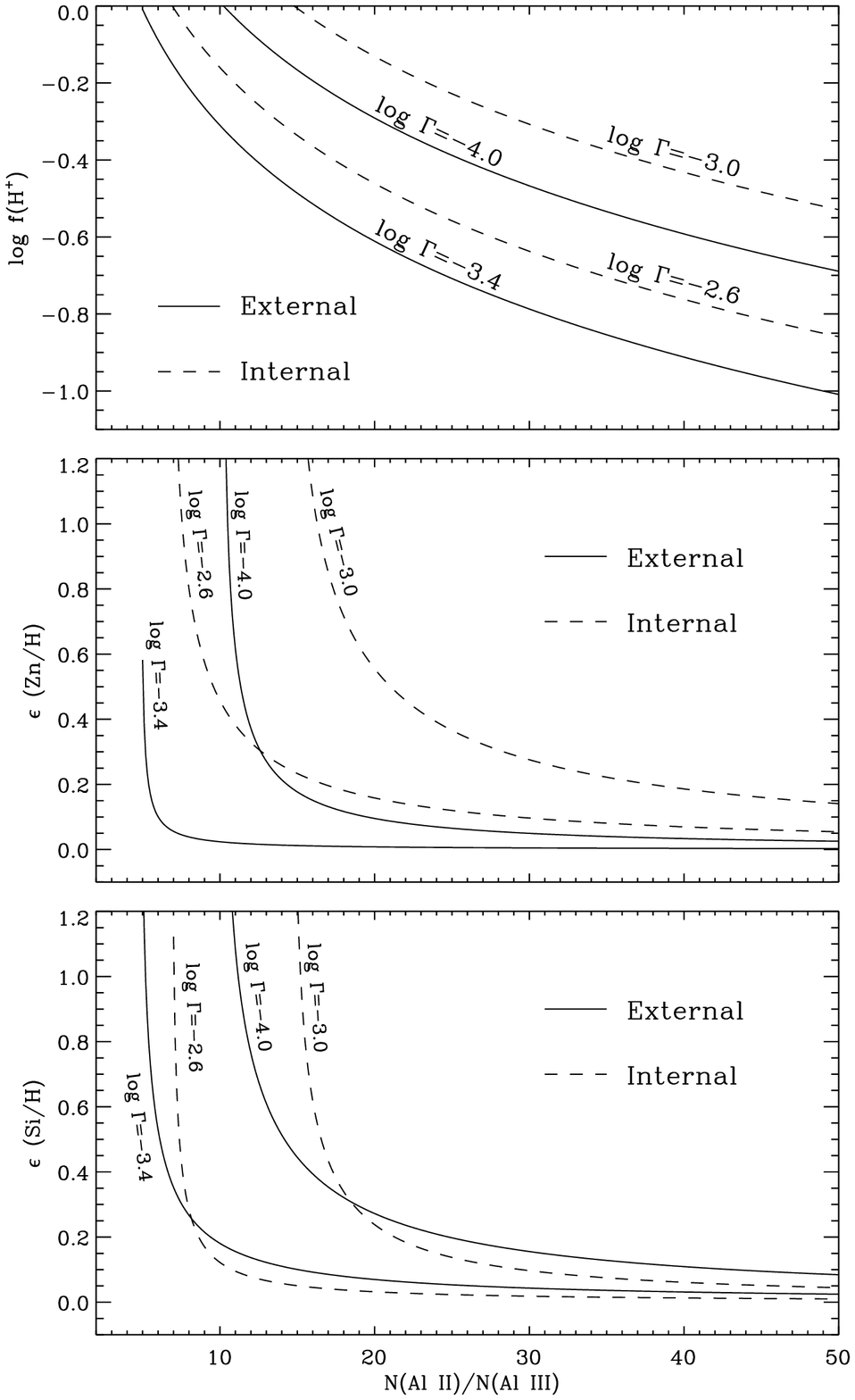}
\caption{The ratio $N(\mbox{\ion{Al}{2}})/N(\mbox{\AlIII})$ 
as an indicator of the ionized gas content of \dlas.  The top panel
shows the fraction of ionized material, $f({\rm H^+})$, in \dlas\ as a
function of the measured $N(\mbox{\ion{Al}{2}})/N(\mbox{\AlIII})$
assuming the labelled ionization parameters.  The bottom two panels
show the error (in dex) of the implied [Zn/H] and [Si/H] abundances if
no correction for ionization is made (see text).  The $\epsilon({\rm
Zn/H})$ results are mostly schematic given the uncertainties in the
atomic data for zinc.
\label{fig:hydrogen}}
\end{figure}

Figures \ref{fig:hmfracs} and \ref{fig:stellarfracs} show that even in
the case where the \ion{Al}{2}/\AlIII\ ratio is large, the amount of
ionized gas in \dlas\ can be significant.  Comparing certain metal
ions to hydrogen may very well systematically {\em overestimate} the
abundances of \dlas.  Figure \ref{fig:hydrogen} shows the implied
fraction of ionized hydrogen in \dlas, \fhii, for the stellar and QSO
ionizing spectra in the top panel, where we have plotted the results
for several different values of $\log \Gamma$.  In the middle panel we
show the logarithmic error introduced into measurements of [Zn/H],
defined as
\begin{equation}
\epsilon ({\rm Zn/H}) 
  \equiv \log \frac{N(\mbox{\ion{Zn}{2}})}
	{N(\mbox{\HI})} \Big|_{measured}
        - \log \frac{N({\rm Zn})}{N({\rm H})}\Big|_{intrinsic},
\end{equation}
for changing mixtures of neutral and ionized gas, as traced by the
\ion{Al}{2}/\AlIII\ ratio, while the bottom panel shows the equivalent
$\epsilon({\rm Si/H})$.  The predicted values of $\epsilon({\rm
Zn/H})$ and $\epsilon({\rm Si/H})$ vary significantly with the adopted
ionizing spectrum and ionization parameter.  Errors in the derived
values of [Zn/H] or [Si/H] of a few tenths of a dex are easily
achievable even when $N(\mbox{\ion{Al}{2}}) \gg
N(\mbox{\ion{Al}{3}})$.

It should be pointed out that the atomic data for zinc are quite
uncertain, with the recombination coefficients being derived from
extrapolations of the results for other elements.  The atomic data for
silicon are more reliable, though the abundance of this element is
complicated by its possible inclusion into dust grains.  The behavior
of $\epsilon({\rm Zn/H})$ and $\epsilon({\rm Si/H})$ observed in
Figure \ref{fig:hydrogen} is a common feature for those elements
predominantly found in their singly-ionized stage in neutral gas.
Figure \ref{fig:hydrogen} shows that \dlas\ with $f({\rm H^+}) \approx
(0.5, \ 0.4, \ {\rm and} \ 0.2)$ can have errors of $\epsilon({\rm
Si/H}) \approx (0.1, \ 0.07, \ {\rm and} \ 0.04)$ dex and
$\epsilon({\rm Zn/H}) \approx (0.3, \ 0.2, \ {\rm and} \ 0.1)$ dex in
the case of internal ionization for $\log
\Gamma = -3.0$.  This error is larger for smaller ionization parameters.
For the external ionizing spectrum these values are $\epsilon({\rm
Si/H}) \approx (0.3, \ 0.2, \ {\rm and} \ 0.05)$ dex and
$\epsilon({\rm Zn/H}) \approx (0.1, \ 0.07, \ {\rm and} \ 0.03)$ dex.

The large spread in total metal abundances, [Zn/H] (Pettini \etal\
1997a, 1999), in \dlas\ at a given redshift could in part be due to
differing ionization conditions.  The {\em total} spread in abundance
at a given redshift can be as high as almost 2.0 dex (Pettini \etal\
1997a, 1999), which is not easily explained by ionization effects.
However, the standard deviations of measurements in a given redshift
interval are of order $0.3-0.4$ dex (Pettini
\etal\ 1997a).  This degree of variation is consistent with  a range of 
\fhii\ values between $\sim0.0$ and $\sim0.6$ in these systems.

If ionization is playing a significant role in determining the
apparent distribution of metallicity in \dlas, we might expect lower
column density systems to show higher inferred abundances, on average.
This is consistent with the claim by Pettini \etal\ (1999) that the
``census'' of metals in known \dlas\ is dominated by high column
density, low metallicity systems, while those higher apparent
metallicity systems tend to be of lower neutral hydrogen column
densities (see also Wolfe \& Prochaska 1998).  However, one should
also be wary of the possible selection effects in identifying high
metallicity, high column density absorbers (Pei \& Fall 1995; Wolfe \&
Prochaska 1998; see also Pettini \etal\ 1999).

Systematic errors in the relative metal abundances can also be
significant, depending on the ions compared.  Unfortunately,
systematic errors in excess of 20\% can begin to mimic other effects
such as nucleosynthetic enrichment or dust depletion.  For example, if
the internal stellar ionizing spectrum is appropriate, the errors in
the [Si/Fe] abundances inferred from $N({\rm Si^+})/N({\rm Fe^+})$ can
mimic the preferential inclusion of iron into dust grains, or the
enhancement of $\alpha$-elements over iron.  For $f({\rm H^+})
\approx (0.5, \ 0.4, \ {\rm and} \ 0.2)$, the systematic errors
in [Si/Fe] are $\epsilon({\rm Si/Fe})\approx (+0.4,
\ +0.3, \ {\rm and} \ +0.2)$.  Similarly, systematic errors 
in the [Cr/Zn] abundances can mimic the inclusion of chromium into
dust: $\epsilon({\rm Cr/Zn})\approx (-0.3, \ -0.2, \ {\rm and}
\ -0.1)$ for the same \fhii\ values. The values \fhii\ required to explain 
the dispersion in inferred [Zn/H] metallicities are also sufficient to
provide the dispersion in inferred [Cr/Zn] values (Pettini \etal\
1997b).

There are some ionic ratios that are accurate tracers of relative
metal abundances even if ionized gas makes a substantial
contribution. For $f({\rm H^+}) < 0.5$, the ratios of \ion{Mn}{2} and
\ion{Mg}{2} to \ion{Fe}{2} should trace Mn/Fe and Mg/Fe to within
$\sim10\%$ in the case of the external (hard) ionizing spectrum.  The
ratio of \ion{Si}{2} to \ion{Al}{2} should be a reasonable proxy for
Si/Al.  For the softer stellar spectrum, the ratios of \ion{Ni}{2} and
\ion{Mg}{2} to \ion{Si}{2} are reliable tracers of Ni/Si and Mg/Si.

\ion{Fe}{3} is a much better tracer of ionized gas than
\ion{Al}{3} in the sense that it is the dominant ionization 
stage of iron in the photoionized gas.  The \wave{1122} transition of
\ion{Fe}{3} may be lost in the \lya\ forest toward high-redshift quasars, 
but in select cases this important transition may be useful for
providing further information on the ionized gas in the \dlas.

Our calculations suggest that ionized regions may make a significant
contribution to the total column density of metal ions in
\dlas, and that this contribution can lead to systematic errors in the
determination of abundances in these systems.  Observational studies
of abundances in \dlas\ should take ionization into account whenever
possible, or at the very least assess its possible impact on the
derived results.


\acknowledgements

We thank G. Ferland and collaborators for their years of work on the
\cloudy\ ionization code, and  F. Haardt and P. Madau for
providing us an electronic version of their updated UV background
spectrum.  Our thanks also to M. Pettini, J. Lauroesch, and
J. Prochaska for helpful comments that have improved the presentation
of our work.  We acknowledge support from the NASA LTSA grant
NAG5-3485.




\begin{references}

\reference{} Baldwin, J., Ferland, G.J., Martin, P.G., Corbin, M., Cota, S., 
Peterson, B.M., \& Slettebak, A., 1991, ApJ, 374, 580

\reference{} Bland-Hawthorn, J. \& Maloney, P.R.  1999, ApJ, 510, L33

\reference{} Bunker, A.J., Warren, S.J., Clements, D.L., Williger, G.M., 
\& Hewett, P.C. 1999, \mnras, in press (astro-ph/9906175)

\reference{} Diplas, A., \& Savage, B.D. 1994, \apjs, 93, 211

\reference{} Fedchak, J.A., \& Lawler, J.E. 1999, ApJ, submitted.

\reference{} Ferland, G.J. 1996, Hazy, a Brief Introduction to CLOUDY 90, 
University of Kentucky Department of Physics and Astronomy Internal
Report

\reference{} Ferland, G.J., Korista, K.T., Verner, D.A., Ferguson,
J.W., Kingdon, J.B., \& Verner, E.M. 1998, \pasp, 110, 761

\reference{} Haardt, F., \& Madau, P. 1996, \apj, 461, 20 (HM)

\reference{} Haffner, L.M., Reynolds, R.J., \& Tufte, S.L. 1999, \apj, 
in press (astro-ph/9904143)

\reference{} Howk, J.C., \& Savage, B.D. 1999, ApJ, 517, 746

\reference{} Howk, J.C., Savage, B.D., \& Fabian, D.  1999, ApJ, in press 
(astro-ph/9905187)

\reference{} Kurucz, R.L. 1991, in Proceedings of the Workshop on
Precision Photometry: Astrophysics of the Galaxy, ed. A.C. Davis
Philip, A.R. Upgren, \& K.A. James (Schenectady: Davis), p. 27.

\reference{} Lauroesch, J.T., Truran, J.W., Welty, D.E., \& York, D.G.
 1996, \pasp, 108, 641

\reference{} Lowenthal, J.D., Hogan, C.J., Green, R.F., Woodgate, B.,
Caulet, A., Brown, L., \& Bechtold, J. 1995, \apj, 451, 484

\reference{} Lu, L., Sargent, W.L.W., Barlow, T.A., Churchill, C.W.,
\& Vogt, S. 1996, \apjs, 107, 475

\reference{} Madau, P., Haardt, F., \& Rees, M.J. 1999, \apj, 514, 648

\reference{} Madau, P., \& Shull, J.M. 1996, \apj, 457, 551

\reference{} McKee, C.F. 1989, \apj, 345, 782

\reference{} McKee, C.F., \& Williams, J.P. 1997, \apj, 476, 144

\reference{} Mezger, P.G. 1987, in Starbursts and Galaxy Evolution,
ed. T.X. Thuah, T. Montmerle, and J. Tran Thanh Van (Paris: Editions
Fronti\`{e}res), p. 3.

\reference{} Pei, Y.C., \& Fall, S.M. 1995, \apj, 454, 69

\reference{} Pettini, M., Ellison, S.L., Steidel, C.C., \& Bowen, D.V. 
1999, \apj, 510, 576

\reference{} Pettini, M., King, D., Smith, L.J., \& Hunstead, R.W. 
1997b, \apj, 478, 536

\reference{} Pettini, M., Smith, L.J., King, D.L., \& Hunstead, R.W.
1997a, \apj, 486, 665

\reference{} Prochaska, J.X. 1999, \apj, 511, L71

\reference{} Prochaska, J.X., \& Wolfe, A.M. 1999, \apjs, 121, 369

\reference{} Reynolds, R.J. 1993, in Back to the Galaxy, 
ed. S. Holt \& F. Verter (New York: American Institute of Physics),
p. 156

\reference{} Reynolds, R.J., Hausen, N.R., Tufte, S.L., \& Haffner, L.M. 1998,
\apj, 494, L99

\reference{} Reynolds, R.J., \& Tufte, S.L. 1995, \apj, 439, L17

\reference{} Sembach, K.R., Howk, J.C., Ryans, R.S.I., Keenan, F.P.
1999, \apj, in press.

\reference{} Savage, B.D., Edgar, R.J., \& Diplas, A. 1990, \apj, 361, 107


\reference{} Sofia, U.J., \& Jenkins, E.B. 1998, \apj, 499, 951

\reference{} Songaila, A., \& Cowie, L.L. 1996, \aj, 112, 335

\reference{} Spitzer, L. \& Fitzpatrick, E.L.  1993, ApJ, 409, 299


\reference{} Steidel, C.C., \& Sargent, W.L.W. 1989, \apj, 343, L33

\reference{} Warren, S.J., \& M{\o}ller, P. 1996, \aap, 311, 25

\reference{} Wolfe, A.M., \& Prochaska, J.X. 1998, \apj, 494, L15

\end{references}
\end{document}